\documentclass[11pt]{article} 
\usepackage{amsfonts,amssymb,slashed,makeidx,latexsym,setspace}
\usepackage{graphicx,graphics,floatflt,amssymb,epsf,rotate,subfigure} 
\usepackage{times,axodraw,cite,color}
\def\ltap{\raisebox{-.4ex}{\rlap{$\sim$}} \raisebox{.4ex}{$<$}}
\def\gtap{\raisebox{-.4ex}{\rlap{$\sim$}} \raisebox{.4ex}{$>$}}

\begin{document}
\begin{flushright}
SINP/TNP/2010/03
\end{flushright}

\vskip 30pt

\begin{center}
{\Large \bf A phenomenological study of 5d supersymmetry} \\
\vspace*{1cm} \renewcommand{\thefootnote}{\fnsymbol{footnote}} { {\sf Gautam
    Bhattacharyya${}$} and {\sf Tirtha Sankar Ray $ {}$} } \\
\vspace{10pt} {\small ${}$ {\em Saha Institute of Nuclear Physics, 1/AF Bidhan
    Nagar, Kolkata 700064, India }}
\normalsize
\end{center}

\begin{abstract}
  Supersymmetry and extra dimension need not be mutually exclusive options of
  physics for the TeV scale and beyond. Intrinsically higher dimensional
  top-down scenarios, e.g. string models, often contain supersymmetry at the
  weak scale. In this paper, we envisage a more phenomenological scenario by
  embedding the 4d constrained minimal supersymmetric standard model in a flat
  5d $S_1 /Z_2$ orbifold, with the inverse radius of compactification at the
  TeV scale. The gauge and Higgs supermultiplets are placed in the bulk. We
  assume that only the third generation matter multiplet accesses the bulk,
  while the first two generations are confined to a brane on an orbifold fixed
  point. From a 4d perspective, the bulk has $N=2$ supersymmetry which entails
  a special non-renormalization theorem giving rise to a significant numerical
  impact on the renormalization group running of various parameters.  The
  brane supersymmetry corresponds to $N=1$ which we assume to be broken in an
  unspecified but phenomenologically acceptable way. Given this set-up, we
  study how the gauge and Yukawa couplings and the $N=1$ brane supersymmetry
  breaking soft masses run through the energy scale exciting the Kaluza-Klein
  states at regular interval. In the process, we ensure that electroweak
  symmetry does break radiatively.  We confront our low energy parameters with
  the experimental measurements or limits of different observables, e.g. LEP
  lower limits on the lightest Higgs boson and the chargino, the $(g-2)$ of
  muon, the branching ratio of $b \to s \gamma$, and the WMAP probe of
  relative dark matter abundance.  We present our results by showing the
  allowed/disallowed zone in the plane of the common scalar mass ($m_0$) and
  common gaugino mass ($M_{1/2}$) for both positive and negative $\mu$
  parameter. Our plots are the first 5d versions of the often displayed 4d
  $m_0$--$M_{1/2}$ plots, and we provide reasons behind the differences
  between the 4d and 5d plots.

\vskip 5pt \noindent
\texttt{PACS Nos:~11.30.Pb, 14.80.Rt} \\
\texttt{Key Words:~Supersymmetry, Extra dimension}
\end{abstract}

\setcounter{footnote}{0}
\renewcommand{\thefootnote}{\arabic{footnote}}

\section{Introduction}
The yet elusive Higgs boson is the primary target of the CERN Large Hadron
Collider (LHC). But, LHC is also expected to reveal a new ruler of the
tera-electron-volt (TeV) territories. The standard model (SM) has so far been
remarkably successful in explaining physics up to a few hundred GeV energy
scale. But theoretical inconsistencies of the SM (like, gauge hierarchy
problem) and experimental requirements (like, a candidate to account for the
dark matter of the universe) suggest that there are good reasons to believe
that new physics beyond the SM is just around the corner crying out for
verification. Among the different possibilities, supersymmetry and extra
dimension stand out as the two leading candidates for dictating terms in the
TeV regime. These two apparently distinct classes of scenarios cover a wide
variety of more specific models. The usual practice from a bottom-up approach
is to attach an `either/or' tag on supersymmetry and extra dimension, as if
the presence of one excludes the other. A more careful thought would reveal
that the relationship between these two is {\em not necessarily} mutually
exclusive.  In fact, the presence of higher dimensions is a common feature of
any fundamental theory valid at high scale.  We will get back to this issue a
little later. For the moment, to put things into perspective, we recapitulate
the chronological evolution of the extra dimensional scenarios without
invoking supersymmetry {\em a priori}.  We restrict our discussion to the flat
space scenarios, as we are not pursuing the warped path in this paper besides
mentioning it as an aside.

Flat extra dimensions were first studied \cite{add} in a scenario where
gravity propagates in a millimeter (mm) size compact space dimension, with the
SM particles confined to a 4d brane. The motive was to bring down the
fundamental Planck scale to about a TeV. Subsequently, it was conceived that
the brane where the SM particles live may actually have a very small size,
like $10^{-16}$ cm $\sim {\rm TeV}^{-1}$, leading to the concept of a `fat
brane' \cite{Antoniadis:1990ew}. In the context of the present paper, we stick
to the fat brane scenario. {\em What are the experimental bounds on the
  fatness of such a brane, more precisely, on the radius of compactification
  ($R$)}? For universal extra dimension (UED) models \cite{acd}, in which {\em
  all} the SM particles access the extra dimensional bulk, a safe estimate is
$R^{-1} ~\gtap~ 500$ GeV.  More specifically, the $g-2$ of the muon
\cite{nath}, flavor changing neutral currents \cite{chk,buras,desh}, $Z \to
b\bar{b}$ decay \cite{santa}, the $\rho$ parameter \cite{acd,appel-yee}, and
hadron collider studies \cite{collued} reveal that $R^{-1}~\gtap~300$
GeV. Consideration of $b \to s \gamma$, however, implies a somewhat tighter
bound ($R^{-1}~\gtap~600$ GeV \cite{bsg}).  Methods to decipher its signals
from the LHC data have recently been discussed too
\cite{Bhattacharyya:2009br}.  On the other hand, in the non-universal scenario
where both the SM gauge bosons and the Higgs boson propagate in the bulk but
the fermions are confined to a 4d brane \cite{nued}, $R^{-1}$ cannot be below
$(1-2)$ TeV \cite{nued-bounds}. The reason behind the difference in
constraints is the following.  The KK parity, defined by $(-1)^n$ for the
$n$th KK label, is conserved in UED, while it is not a good symmetry in the
non-universal scenario. As a result, while in the non-universal models KK
states can mediate many processes at tree level yielding strong constraints,
in the UED model, thanks to the KK parity, KK states appear only inside a loop
leading to milder constraints.  {\em In any case, in the presence of
  supersymmetry, all those analyses need to be modified with more parameters,
  which would expectedly lead to a set of more relaxed bounds on $R^{-1}$}.

{\em Now, what is the motivation of studying a TeV scale (or, a fat brane)
  extra dimension scenario}? The implications of such models have been
investigated from the perspective of string theory, phenomenology,
cosmology/astrophysics and high energy experiments.  Such models provide a
cosmologically stable dark matter candidate \cite{Servant:2002aq}, trigger
electroweak symmetry breaking successfully through a composite Higgs
\cite{Arkani-Hamed:2000hv}, address the fermion mass hierarchy problem from a
different point of view \cite{Arkani-Hamed:1999dc}\footnote{Generation
  non-universality in fermion localization imposes $R^{-1} > 5000$ TeV due to
  large flavor-changing neutral currents and CP violation
  \cite{Delgado:1999sv}.}, and stimulate power law renormalization group (RG)
running yielding a lower (few tens of a TeV) gauge coupling (near-)unification
scale \cite{Dienes:1998vg,dienes2,blitz}\footnote{The power law loop
  corrections are admittedly ultraviolet (UV) cutoff dominated. It has been
  argued that if the higher dimensional theory contains a larger gauge
  symmetry which is perturbatively broken then the difference of gauge
  couplings of the unbroken subgroups is a calculable quantity independent of
  UV completion \cite{Hebecker:2002vm}.}. Besides, the running of neutrino
mixing angles generated from effective Majorana mass operator in a 5d set-up
has been studied both in non-supersymmetric \cite{Bhattacharyya:2002nc} and
supersymmetric \cite{Deandrea:2006mh} contexts.

{\em But, what is the advantage of supersymmetrizing it?} We argue that
supersymmetry and extra dimension need not always be seen as {\em two} new
physics considered simultaneously. In fact, they may nicely complement each
other in {\em some situations} through mutual requirements. This can be seen
as follows:
\begin{enumerate} 
\item From a top-down approach, string theory provides a rationale behind
  linking supersymmetry and extra dimension. The string models are
  intrinsically extra dimensional, and more often than not contain
  supersymmetry as an integral part.  That said, we must also admit that
  establishing a rigorous connection between a {\em realistic} low energy
  supersymmetric model with string theory is still a long shot, though a lot
  of efforts have already been put in that direction \cite{Acharya:2007rc}.

\item The higher dimensional field theories are non-renormalizable, and
  computation of quantum corrections in such scenarios require the existence
  of a UV completion. Superstring theories, which contain supersymmetry,
  provide the main hope in this context. In fact, a joint r\^ole of `partial
  supersymmetry' (broken at very high scale) and warped extra
  dimension/compositeness has been envisaged from AdS/CFT point of view to
  account for both the `little' and the `big' hierarchies
  \cite{Sundrum:2009gv}.

\item Even after embedding the SM in an extra dimensional set-up, the scalar
  potential remains unstable under quantum correction. Supersymmetrization
  stabilizes it and ameliorates the hierarchy problem.  It is interesting to
  note that by admitting chiral fermions and their scalar partners in the same
  multiplet tacitly provides a rationale behind treating the Higgs boson as an
  elementary object. An elementary Higgs can be perfectly accommodated in a
  flat extra dimensional set-up. As a corollary, the upper limit on the
  lightest supersymmetric neutral Higgs is relaxed beyond the 4d upper limit
  of 135 GeV due to the presence of the KK towers of top/stop chiral
  multiplets, and the {\em hitherto} disfavored low $\tan\beta$ region can be
  resurrected \cite{Bhattacharyya:2007te}.

\item Each 4d supersymmetric scenario has its own supersymmetry breaking
  mechanism. The origin of this mechanism may be linked to the existence of
  extra dimension. In fact, one of the earliest motivations of a TeV scale fat
  brane was to relate the scale of 4d supersymmetry breaking with $R^{-1}$
  \cite{Antoniadis:1990ew,Antoniadis:1992fh}.

\end{enumerate} 

Keeping these in mind, we outline the formalism of a 5d supersymmetric model
in an $S^1/Z_2$ orbifold which contains the 4d supersymmetric states as zero
modes. In section \ref{5dmssm}, we state our assumptions leading to the
construction of the 5d model and comment on supersymmetry
breaking. Furthermore, we explicitly write down the particle content and their
5d Lagrangian and illustrate the KK decompositions of the different 5d
fields. In section \ref{beta}, we derive the beta functions of the gauge and
Yukawa couplings as well as those of the different soft supersymmetry breaking
parameters {\em diagram by diagram}, pointing out how they are all modified
from their 4d values due to the presence of KK states. In section
\ref{numerical}, we discuss the numerical effects of RG running and highlight
the reason behind the differences between the 4d and 5d scenarios. We also
point out under what conditions we can ensure electroweak symmetry
breaking. In section \ref{m0mhalf}, we exhibit the numerical impact of RG
running through plots showing constraints in the $m_0$--$M_{1/2}$ plane. We
standardize our numerical codes by reproducing the known 4d plots before
encoding the necessary alterations for producing the new plots pertaining to
5d supersymmetry.  This also enables us to compare and contrast the 4d and 5d
allowed/disallowed zones. Finally, in section \ref{concl}, we showcase the
essential features we have learnt from this analysis.

\section{5d supersymmetry}
\label{5dmssm} 
\subsection{A snapshot of our model}
We highlight the salient features of supersymmetry in higher dimension and
outline below the various assumptions that lead to a calculable
phenomenological framework.
\begin{enumerate}
	
\item We consider a 5d flat space time metric. The 5th dimension is
  compactified on a $S_1/Z_2$ orbifold. Orbifolding is necessary to reproduce
  chiral zero mode fermions as a 5d theory is vector-like. 
	
\item We embed the minimal supersymmetric standard model (MSSM) in this higher
  dimensional set-up (several consequences of such embedding, mainly the
  effects on gauge and Yukawa couplings' evolution, have been studied in
  \cite{Dienes:1998vg}). From a 4d point of view this leads to a tower of KK
  states. The massless sector corresponds to the 4d MSSM states.
        
\item Since in 5d bulk the fermion representation is vectorial, the
  two-component spinor $Q$ that generates 4d supersymmetry will in 5d be
  accompanied by its chiral conjugate mirror $Q^c$. Thus a $N=1$ supersymmetry
  in 5d corresponds to two different $N=1$ supersymmetry, or equivalently, a
  $N=2$ supersymmetry from a 4d perspective. In fact, all members of a given
  KK mode must fall into a valid representation of $N=2$ supersymmetry.  In
  fact, each 4d supermultiplet is augmented by new chiral conjugate states and
  together they form a hypermultiplet.

\item Here we are talking about a massive representation of supersymmetry,
  where the supersymmetry preserving Dirac mass plays the r\^ole of central
  charge for $N=2$ supersymmetry. This charge is not renormalized, as a
  consequence of which the bulk hypermultiplets do not receive any
  wave-function renormalization\footnote{In other words, for $N=2$
    supersymmetry it turns out that $m_R= m_B$, which is analogous to $g_R =
    g_B$ for $N=4$ supersymmetry. Here $m$ is the Dirac mass (central charge)
    and $g$ is gauge coupling, while $R$ and $B$ are labels for renormalized
    and bare quantities. Since the Dirac mass of $N=2$ hypermultiplets appears
    on the right-hand side of the anti-commutation relation of the conserved
    supersymmetry charges, this mass cannot be renormalized. This is
    intertwined with the observation that only those terms are renormalized
    which can be written as integrals over all superspace volume. The kinetic
    term of $N=2$ hypermultiplets cannot be written as any such integral (see
    discussions and related earlier references in \cite{Barbieri:1982nz}.}
  \cite{Dienes:1998vg,Barbieri:1982nz}. We observe that this $N=2$
  non-renormalization has serious numerical consequences in RG evolution of
  parameters. The most notable effect is the blowing up of the Yukawa
  couplings into the non-perturbative regime around 18 TeV, which we will take
  to be the cutoff of our theory. This is below the scale of perturbative
  gauge coupling unification, which is around 30 TeV. Recall that in 5d we
  encounter power law running which results in early (compared to 4d)
  unification.

\item We allow the gauge and the Higgs multiplets access the 5d bulk. Thus far
  what we said is nothing but a supersymmetrization of UED. Only the matter
  multiplets make the difference. In the UED framework, {\em all} SM particles
  access the bulk, and thus even though there are two fixed points, there is
  no brane. One could as well have kept some or all of the fermion generations
  in a brane at a fixed point; the difference would be that the scenario would
  cease to be universal. In the present supersymmetric context too we have the
  freedom of keeping some or all of the matter multiplets at an orbifold fixed
  point. We note that unless we confine at least two generations of matter
  multiplets on a brane, the requirement of {\em perturbative} gauge coupling
  unification leads to a constraint $R^{-1} > 10^{10}$ GeV \cite{blitz},
  spoiling its relevance for LHC. On the other hand, unless we keep the third
  family of matter multiplet in the bulk we cannot ensure electroweak
  breaking. In view of the above, we let the third generation matter multiplet
  access the bulk, but fix the first two generations at the $y=0$ brane.

\item $N=2$ supersymmetry forbids Yukawa interaction in the 5d bulk as this
  interaction involves odd (three) number of {\em chiral}
  multiplets. Therefore, we localize Yukawa interaction at the orbifold fixed
  point where the supersymmetry corresponds to $N=1$.

\item Now we come to the important question as how we break the residual $N=1$
  supersymmetry. Different ideas have been advanced for its realization. One
  way is to break it by the Scherk-Schwarz mechanism \cite{ss} in which
  fermions and bosons satisfy different periodic conditions over the
  compactified dimension. Explicit realizations towards this using a TeV-scale
  orbifold can be found in \cite{pomarol}. Another interesting approach was to
  break the residual supersymmetry by a second compactification on an orbifold
  with two reflection symmetries, viz.~$S^1/(Z_2 \times Z_2^\prime)$
  \cite{Barbieri:2000vh}. This can be viewed as a discrete version of the
  Scherk-Schwarz mechanism. Both these scenarios yield soft masses which are
  UV insensitive due to the non-local nature of supersymmetry breaking. From a
  completely different viewpoint, supersymmetry breaking may be infused from
  the brane-bulk interface \cite{Mirabelli:1997aj}, or transmitted from a
  distant brane \cite{Kaplan:1999ac}, or arisen from a gaugino mediation
  set-up \cite{Chacko:1999mi} (possibly with a much lower cutoff than
  $10^{16}$ GeV), or triggered by some completely unknown brane dynamics, for
  example, by a spurion $F$-term vacuum expectation value (vev).

  In the context of the present analysis, we keep the exact mechanism of the
  $N=1$ brane supersymmetry breaking {\em unspecified}.  We assume that the
  supersymmetry breaking scale is of the order of the inverse radius of
  compactification, for example $c/R$, where $c$ is an ${\cal {O}}(1)$
  dimensionless parameter.  

\item {\em Our main goal is the following}: Just like in the conventional but
  constrained version of 4d supersymmetry one starts with a common scalar and
  a common gaugino mass at high scale (e.g. the GUT scale) and then run them
  down using the MSSM beta functions to find the weak scale spectrum, we do
  exactly the same here by assuming a common scalar mass ($m_0$) and a common
  gaugino mass ($M_{1/2}$) at low cutoff scale (18 TeV) and follow the running
  using the KK beta functions through successive KK thresholds to obtain the
  weak scale parameters. By adopting a phenomenological approach, we scan
  $m_0$ and $M_{1/2}$ over a set of values $c/R$, with $c$ varying in the
  range $0.1$ to $1$ and $R^{-1}$ fixed at $1$ TeV.

\end{enumerate}
 
\subsection{Multiplet Structures}
As mentioned in the Introduction, from a 4d perspective the KK towers of
matter and gauge fields rearrange in the form of $N=2$ hypermultiplets. A
judicious choice of $Z_2$ parity of the 5d fields allows us to break the $N=2$
supersymmetry to $N=1$ supersymmetry.  We briefly review below the multiplet
structures of the fields following the prescription suggested in
\cite{ArkaniHamed:2001tb}.

\subsubsection{Vector hypermultiplet}
The 5d super Yang-Mills theory contains a 5-vector gauge field, a 4-component
Dirac gaugino and a real scalar. When dimensionally reduced to 4d, the gauge
field splits into a 4-vector and a scalar, the gaugino splits into 2 Majorana
gauginos, and we still have the real scalar previously mentioned. All these
fit into a vector multiplet and a chiral multiplet in $N=1$ language.  If we
represent the $N=2$ vector supermultiplet by $V$, the 4-vector gauge field by
$A_{\mu}$, the gauginos by $\lambda$ and $\psi$, and define a complex scalar
field $\phi\equiv \frac{1}{2}(\Sigma + iA_5)$, where $\Sigma$ is the 5d real
scalar and $A_5$ is the 5th component of the 5-vector field, then one can
schematically represent the 5d vector supermultiplet as
\begin{equation}
\label{5dV}
  V \equiv \pmatrix{ A_{\mu} & \phi\cr \lambda & \psi \cr} \, . 
\label{Tvec}
\end{equation}
From a 4d perspective (where the compactified 5th coordinate $y$ is just a
label), and in the $N=1$ language, one can visualize the vector hypermultiplet
by a vector multiplet $\mathcal{V}$ (first column) and a chiral multiplet in
the adjoint representation by $\Phi$ (second column):
\begin{eqnarray}
{\mathcal{V}}(x,y) &=& -\theta {\sigma}^{\mu} \overline{\theta} A_{\mu}{(x,y)}
+ i ~{\overline{\theta}}^2 \theta {\lambda}{(x,y)} - i~ {\theta}^2
\overline{\theta}~{\overline{\lambda}}{(x,y)} + \frac{1}{2}
 {\overline{\theta}}^2 {\theta}^2 D_V(x,y) \, , \nonumber \\ 
{\Phi}{(x,y)} &=& {\phi}{(x,y)} + \sqrt{2}~ \theta {\psi}{(x,y)} + 
{\theta}^2 F_{\Phi}(x,y)  \, .
\end{eqnarray}
The $Z_2$ parity of $V$ is so chosen that the $\mathcal{V}$ contains a zero
mode, but $\Phi$ does not have any zero mode.

The gauge invariant action may be written as ($\int d^5x \equiv \int d^4x \int
dy$)
\begin{equation}
 S^5_{\rm gauge} = \int d^5 x \left[ \frac{1}{4 g^2} \int d^2 \theta
   ~W^{\alpha} W_{\alpha} + {\rm h.c.} + \int d^4 \theta ~\frac{1}{g^2}
   {\left( {\partial}_5 \mathcal{V} - \frac{1}{\sqrt{2}} \left( \Phi +
     \overline{\Phi} \right) \right)}^2 \right],
\end{equation} 
where the $~W^{\alpha}(x,y)$ is the field strength superfield
corresponding to ${\mathcal{V}}(x,y)$.

\subsubsection{Higgs hypermultiplets}
From the $N=1$ perspective, the $N=2$ hypermultiplet splits into two chiral
multiplets. Thus have a $H_u$ hypermultiplet and a $H_d$ hypermultiplet.
We can represent them as (the tilde symbol represents Higgsino) 
\begin{equation}
\label{huhd-hyper}
{H_{(u,d)}} \equiv \pmatrix{ {H_{L(u,d)}} & {H_{R(u,d)}}\cr {\tilde{H}_{L(u/d)}}
  & {\tilde{H}_{R(u/d)}} \cr} \, . 
\end{equation} 
If we denote the two chiral multiplets inside the hypermultiplet $H(x,y)$ as
$h(x,y)$ in left column and $h^c(x,y)$ in right column, then one can expand
the chiral superfield as
\begin{eqnarray} {h/h^c} = {H_{L/R}} + \sqrt{2}~ \theta {\tilde{H}_{L,R}} +
  {\theta}^2 F_{h/h^c} \, .
\end{eqnarray}
We assign even $Z_2$ parity to $h$ so that it has a zero mode, and odd $Z_2$
parity to $h^c$ which does not have zero mode. 
The free action of the hypermultiplets 
can be written as
\begin{equation}
S^5_{\rm Higgs} = \int d^5 x \left[ \int d^4 \theta \left(
  {\overline{h}}^{c} h^{c} + {\overline{h}} h \right)
  + \left( \int d^2 \theta ~h^{c} \left( {\partial}_5 + m \right)h
  + {\rm h.c.}\right) \right] \, . 
\label{higgact}
\end{equation}

\subsubsection{Matter hypermultiplets}
Matters have similar hypermultiplet structures similar to Higgs:
\begin{equation}
  \Psi \equiv \pmatrix{
     {{\phi}_L} & {{\phi}_R}\cr
    {{\psi}_L} & {{\psi}_R} \cr}~,
\end{equation}
where, ${{\mathcal{F}}_L} \equiv \left( {{\phi}_L} , {{\psi}_L} \right)$
($Z_2$ even) and ${{\mathcal{F}}_R} \equiv \left( {{\phi}_R} , {{\psi}_R}
\right)$ ($Z_2$ odd) represent the two $N=1$ chiral multiplets. 
The free matter hypermultiplet action will be similar to Eq.~(\ref{higgact}).
There are five matter representations, two SU(2) doublets $Q$ and $L$ and
three singlets $u, d, e$, where the symbols have their standard meaning.

\subsubsection{Gauge interactions}  
When the hypermultiplets are charged under gauge symmetry, their free action 
can be promoted to take care of the interaction in the following way: 
\begin{eqnarray}
S^5_{\rm int} &=& \int d^5x \left[\int d^4 \theta \left(
  {{\mathcal{F}}_L} e^{\mathcal{V}}
  {{\overline{\mathcal{F}}}_L} + {{\mathcal{F}}_R}
  e^{-{\mathcal{V}}} {{\overline{\mathcal{F}}}_R} \right)
+ \left\{ \int d^2\theta {\mathcal{F}}_L
  \left( m + {\partial}_5 - \frac{1}{\sqrt{2}} {\Phi}
  \right) {{\mathcal{F}}_R} + {\rm h.c.} \right\} \right],
\end{eqnarray}
where, ${\mathcal{V}} = {\mathcal{V}}^{a} T^a$ and ${\Phi} =
{\Phi}^{a} T^a$ are Lie-algebra-valued gauge and matter superfields. 

\subsubsection{Yukawa Interactions}
We denote the Yukawa part of the superpotential by $W_Y$, which contains the
usual chiral superfield combinations $QH_uu$, $QH_dd$ and $LH_de$.  Since an
Yukawa interaction involves three (i.e. odd number) chiral superfields, it is
not possible to write a bulk Yukawa interaction maintaining $N=2$
supersymmetry. For this reason, we confine Yukawa interaction at the branes
even though the associated superfields are in the bulk.  A generic action
(e.g. involving the third generation bulk matter) can be written as
\begin{equation}
S^5_{\rm Yuk}= \int d^5x\left( \int d^2 \theta ~
W_Y \right)\left[ \delta(y)+\delta(y-{\pi}R)\right] \, . 
\end{equation} 
As the $Z_2$ odd fields vanish at the fixed points, they do not contribute to
Yukawa interactions. We note that the first two matter generations can be kept
either at either of the two fixed points. Also, the above Yukawa action could
have been confined at either of the two branes instead of symmetrically on
both.

\subsection{KK decomposition of fields}
In order to obtain the action in terms of 4d component fields, we need to
write down the KK decomposition of the 5d fields in terms of zero modes and
higher KK modes \cite{acd}. Each 5d field is either $Z_2$ even or $Z_2$
odd. Only the even fields have zero modes. The decomposition of the Higgs
fields will be exactly like the matter fields. 
\begin{eqnarray}
\label{fourier}
\mathcal{V}(x,y)&=&\frac{\sqrt{2}}{\sqrt{2\pi
R}}\mathcal{V}^{(0)}(x)+\frac{2}{\sqrt{2\pi
R}}\sum^{\infty}_{n=1}\mathcal{V}^{(n)}(x)\cos\frac{ny}{R} \, , \nonumber\\
\Phi(x,y) &=& \frac{2}{\sqrt{2\pi
R}}\sum^{\infty}_{n=1}\Phi^{(n)}(x)\sin\frac{ny}{R} \, , \\
{\mathcal{F}}_L(x,y)&=&\frac{\sqrt{2}}{\sqrt{2\pi
R}}{\mathcal{F}}_L^{(0)}(x)+\frac{2}{\sqrt{2\pi
R}}\sum^{\infty}_{n=1}{\mathcal{F}}_L^{(n)}(x)\cos\frac{ny}{R} \,, \nonumber \\
{\mathcal{F}}_R(x,y)&=&\frac{2}{\sqrt{2\pi
R}}\sum^{\infty}_{n=1}{\mathcal{F}}_R^{(n)}(x)\sin\frac{ny}{R} \, . \nonumber
\end{eqnarray}

\section{RG evolution and derivation of the beta functions}
\label{beta}
The technical meaning of RG evolution in a higher dimensional context has been
amply clarified in \cite{Dienes:1998vg}, and we merely reiterate it in the
present context. The multiplicity of KK states renders any such higher
dimensional scenario non-renormalizable. So `running' of couplings or
parameters with the energy scale does not make much of a sense. Rather, one
can estimate the finite quantum corrections that these couplings/parameters
receive whose size depends on some explicit cutoff $\Lambda$. The contribution
comes from $\Lambda R$ number of KK states which lie between the scale of the
first KK state, which is $1/R$, and the cutoff $\Lambda$. With this
interpretation of RG running, we compute the one loop beta functions of the
gauge and Yukawa couplings and various soft supersymmetry breaking masses.  We
make the following observations:
\begin{enumerate}
\item The contribution to the beta function from a given KK mode does not
  depend on its KK label.
\item When we consider different KK thresholds we neglect their zero mode
  masses, i.e. we assume that the $n$th level KK state is kicked into life
  when we cross the energy scale $n/R$.
\item As we cross different KK thresholds, the beta functions also change.
  The beta function of the quantity $X$ at an energy scale $Q$, where 
  $n~\ltap~QR < (n+1)$, can be written as ($t = \ln (Q/Q_0)$, where $Q_0$ is a
  reference scale, e.g. the electroweak scale)
\begin{equation}
\label{master}
\frac{\partial{X}}{\partial{t}} = \beta_X \, ,~~{\rm where}~~
\beta_X = \beta_{0X} + n \tilde{\beta}_X \, . 
\end{equation}
Here $\beta_{0X}$ is the contribution induced by the zero mode (i.e. ordinary
4d MSSM) states (which may be found, for example, in the review
\cite{Martin:1997ns}) and $\tilde{\beta}_X$ arises from a single KK
mode. Eq.~(\ref{master}) is our master equation using which we perform a
diagram by diagram calculation for the estimation of $\tilde{\beta}_X$ for
various couplings and parameters.
\end{enumerate}

\subsection{Gauge couplings and gaugino masses}
The running of the gauge couplings ($g_i$) and gaugino masses ($M_i$) are
controlled by
\begin{eqnarray}
\label{gaugerg}
\beta_{g_i} = \frac{g_i^3}{16\pi^2} \left[b_i^0 + n \tilde{b}_i\right] \, ,~~  
\beta_{M_i} = \frac{g_i^2 M_i}{16\pi^2} \left[b_i^0 + n \tilde{b}_i \right] \, .
\end{eqnarray}
For the gauge groups U(1) (which cooresponds to $g_1 = \sqrt{5/3} g'$, which
unifies), SU(2) and SU(3), $b_i^0 = (33/5,1,-3)$ \cite{Martin:1997ns}, and
$\tilde{b}_i = (26/5,2,-2)$ \cite{blitz}, respectively.

\subsection{Yukawa and scalar trilinear couplings}
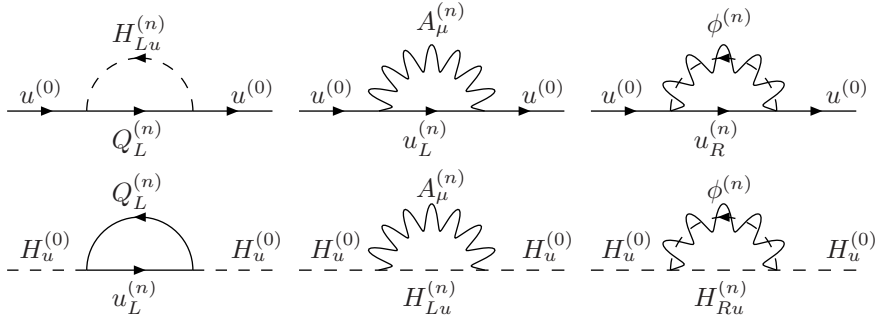
\begin{figure}
\begin{center}
\fcolorbox{white}{white}{
  \begin{picture}(350,92) (105,-89)
    \SetWidth{0.5}
    \Text(110,-25)[lb]{\small{\textbf{$u^{(0)}$ }}}
    \Text(190,-25)[lb]{\small{\textbf{$u^{(0)}$ }}}
    \Text(145,-45)[lb]{\small{\textbf{$Q_{L}^{(n)}$ }}}
    \Text(145,-5)[lb]{\small{\textbf{$H_{Lu}^{(n)}$ }}}
    \Text(220,-25)[lb]{\small{\textbf{$u^{(0)}$ }}}
    \Text(300,-25)[lb]{\small{\textbf{$u^{(0)}$ }}}
    \Text(255,-45)[lb]{\small{\textbf{$u_{L}^{(n)}$ }}}
    \Text(260,0)[lb]{\small{\textbf{$A_{\mu}^{(n)}$ }}}
    \Text(330,-25)[lb]{\small{\textbf{$u^{(0)}$ }}}
    \Text(415,-25)[lb]{\small{\textbf{$u^{(0)}$ }}}
    \Text(365,-45)[lb]{\small{\textbf{$u_{R}^{(n)}$ }}}
    \Text(370,0)[lb]{\small{\textbf{$\phi^{(n)}$ }}}

    \Text(110,-85)[lb]{\small{\textbf{$H_u^{(0)}$ }}}
    \Text(190,-85)[lb]{\small{\textbf{$H_u^{(0)}$ }}}
    \Text(145,-105)[lb]{\small{\textbf{$u_{L}^{(n)}$ }}}
    \Text(145,-65)[lb]{\small{\textbf{$Q_{L}^{(n)}$ }}}
    \Text(220,-85)[lb]{\small{\textbf{$H_u^{(0)}$ }}}
    \Text(300,-85)[lb]{\small{\textbf{$H_u^{(0)}$ }}}
    \Text(255,-105)[lb]{\small{\textbf{$H_{Lu}^{(n)}$ }}}
    \Text(260,-63)[lb]{\small{\textbf{$A_{\mu}^{(n)}$ }}}
    \Text(330,-85)[lb]{\small{\textbf{$H_u^{(0)}$ }}}
    \Text(415,-85)[lb]{\small{\textbf{$H_u^{(0)}$ }}}
    \Text(365,-105)[lb]{\small{\textbf{$H_{Ru}^{(n)}$ }}}
    \Text(370,-63)[lb]{\small{\textbf{$\phi^{(n)}$ }}}
    \ArrowLine(105,-28)(135,-28) \ArrowLine(135,-28)(175,-28) 
    \ArrowLine(175,-28)(205,-28) \DashArrowArc(155,-28)(20,0,180){5}
    \DashLine(105,-88)(135,-88){5} \ArrowLine(135,-88)(175,-88)
    \DashLine(175,-88)(205,-88){5} \ArrowArc(155,-88)(20,0,180)
    \DashLine(215,-88)(315,-88){5} \ArrowLine(215,-28)(245,-28)
    \ArrowLine(245,-28)(285,-28) \ArrowLine(285,-28)(315,-28)
    \PhotonArc(265,-28)(20,0,180){-5}{7.5} \ArrowLine(325,-28)(355,-28)
    \ArrowLine(355,-28)(395,-28) \ArrowLine(395,-28)(425,-28)
    \PhotonArc(375,-28)(20,0,180){-5}{5.5} \DashArrowArc(375,-28)(20,0,180){5}
    \PhotonArc(265,-88)(20,0,180){-5}{7.5} \DashLine(325,-88)(425,-88){5}
    \PhotonArc(375,-88)(20,0,180){-5}{5.5} \DashArrowArc(375,-88)(20,0,180){5}
  \end{picture}
}
\end{center}
\caption{\small{\em Feynman diagrams showing the KK contributions to the
    wave-function renormalizations of the zero mode $u_3$ and $H_u$.  Similar
    diagrams for the other fermions and scalars may be drawn analogously. Here
    $A_{\mu}$ is a generic gauge field and $\phi$ is an adjoint scalar, both
    arising from a vector hypermultiplet, as explained around
    Eq.~(\ref{5dV}).}  }\label{selfenergies}
\end{figure}

We recall that $N=1$ non-renormalization relates the beta functions of the
Yukawa couplings ($y_{ijk}$) to the anomalous dimension matrices
($\gamma^i_j$) of the superfields.  This theorem implies that logarithmically
divergent contributions can always be written in terms of wave-function
renormalizations.  Generically, $y_{ijk}$ may be written as
\begin{equation}
\beta_{y^{ijk}} = \gamma^i_n y^{njk} + \gamma^j_n y^{ink} + \gamma^k_n y^{ijn}
\, . 
\label{nonrenorm}
\end{equation}
The Feynman diagrams showing the KK contributions to the wave-function
renormalizations of the scalars and fermions are displayed in
Fig.~\ref{selfenergies}. The contribution from the gauge sector cancels
exactly as a consequence of the $N=2$ non-renormalization theorem mentioned in
section \ref{5dmssm}.  Diagrammatically, the origin of this cancellation may
be traced to a relative sign between the $A_\mu$- and $\phi$-propagators - see
Eq.~(\ref{5dV}).  Only the brane localized Yukawa interactions contribute to
the Yukawa evolution. We also keep track of the fact that the $Z_2$ odd fields
have vanishing wave-functions at the two branes, leaving the even fields alone
to contribute to the diagrams in Fig.~\ref{selfenergies}. Here we have made
a tacit assumption that although the Yukawa interaction is brane localized,
only one KK level ($n$) states float inside the loop at a time. This is a
technical assumption to ensure calculability by avoiding KK divergence which
would have arisen while summing more than one KK index in a loop calculation.

\normalsize

To appreciate the numerical impact of the bulk $N=2$ non-renormalization, we
first write down the conventional 4d MSSM beta functions (i.e. those coming
from zero mode states in the 5d context) which contribute to the evolution of
the third generation Yukawa couplings \cite{Martin:1997ns}:
\begin{eqnarray}
 \beta_{t}^0 &=& {y_t \over 16 \pi^2} \Bigl [ 6 y_t^* y_t + y_b^* y_b -
   {16\over 3} g_3^2 - 3 g_2^2 - {13\over 15} g_1^2 \Bigr ],\nonumber
 \\ \beta_{b}^0 &=& {y_b \over 16 \pi^2} \Bigl [ 6 y_b^* y_b + y_t^* y_t +
   y_\tau^* y_\tau - {16\over 3} g_3^2 - 3 g_2^2 - {7\over 15} g_1^2 \Bigr] 
\, , \\ \beta_{\tau}^0 &=& {y_\tau \over 16 \pi^2} \Bigl [ 4 y_\tau^*
   y_\tau + 3 y_b^* y_b - 3 g_2^2 - {9\over 5} g_1^2 \Bigr ] \, . \nonumber 
\end{eqnarray}
The corresponding KK contributions are given by 
\begin{equation} 
\label{betaf}
\tilde{\beta}_f = \beta_f^0 (g_i \to 0) ~~ (f \equiv t,b,\tau) \, , 
\end{equation} 
where the vanishing gauge contributions is a direct consequence of the bulk
$N=2$ non-renormalization.

The effects of the above non-renormalization can also be felt in the evolution
of the trilinear scalar couplings.  The relevant Feynman diagrams are
displayed in Fig.~\ref{trilinear}.  Again, for illustration, we first write
down the contributions to the beta functions from the zero mode (i.e. 4d MSSM)
states \cite{Martin:1997ns}:
\begin{eqnarray}
\label{atrge}
{\beta}_{a_t}^0 \!&=&\! \frac{1}{16\pi^2} \Bigl [a_t \left( 18 y_t^* y_t +
  y_b^* y_b - {16\over 3} g_3^2 - 3 g_2^2 - {13\over 15} g_1^2 \right) + 2 a_b
  y_b^* y_t\Bigr .  \nonumber\\ && \!+ ~y_t \Bigl . \left( {32\over 3} g_3^2
  M_3 + 6 g_2^2 M_2 + {26\over 15} g_1^2 M_1 \right) \Bigr ] \, , \nonumber
\\ {\beta}_{a_b}^0 \!&=&\!\frac{1}{16\pi^2}\Bigl [ a_b \left( 18 y_b^* y_b +
  y_t^* y_t + y_\tau^* y_\tau - {16\over 3} g_3^2 - 3 g_2^2 - {7\over 15}
  g_1^2 \right) + 2 a_t y_t^* y_b + 2 a_\tau y_\tau^* y_b \Bigr .
  \phantom{xxxx} \nonumber \\&& \!+ y_b \Bigl. \left( {32\over 3} g_3^2 M_3 +
  6 g_2^2 M_2 + {14 \over 15} g_1^2 M_1 \right) \Bigr ] \, ,\qquad{} 
\\ \beta_{a_{\tau}}^0 \!&=&\! \frac{1}{16\pi^2}\Bigl [a_\tau \left( 12
  y_\tau^* y_\tau + 3 y_b^* y_b - 3 g_2^2 - {9\over 5} g_1^2 \right) + 6 a_b
  y_b^* y_\tau + y_\tau \left( 6 g_2^2 M_2 + {18\over 5} g_1^2 M_1 \right)
  \Bigr ] \, . \nonumber
\end{eqnarray}
As expected, the beta functions of the {\em soft supersymmetry breaking
  parameters} are proportional {\em not only} to those parameters but to
others as well, since any non-renormalization theorem ceases to work when
supersymmetry is broken. For the computation of $\tilde{\beta}_{a_f}$, we need
to keep in mind the essence of Eq.~(\ref{betaf}), i.e. the {\em absence of
  gauge contributions} in $\tilde{\beta}_{f}$, while solving the coupled
differential equations. However, that part of the gauge contributions
(proportional to $g_i^2$) to the trilinear scalar couplings which multiply the
gaugino masses ($M_i$) in Eq.~(\ref{atrge}) would still remain while computing
the KK contribution.  All in all,
\begin{equation} 
\label{betaaf}
\tilde{\beta}_{a_f} = \beta_{a_f}^0 (a_f g_i \to 0) \, . 
\end{equation} 
\begin{center}
 \begin{figure}[t]
\begin{center}
\begin{center}
\fcolorbox{white}{white}{
  \begin{picture}(318,255) (60,-75)
    \SetWidth{0.5}
\Text(65,125)[lb]{\small{\textbf{$H_u^{(0)}$ }}}
\Text(118,134)[lb]{\small{\textbf{$\tilde{u}^{(0)}$ }}}
\Text(170,47)[lb]{\small{\textbf{$\tilde{Q}^{(0)}$ }}}
\Text(150,134)[lb]{\small{\textbf{${Q}_L^{(n)}$ }}}
\Text(128,165)[lb]{\small{\textbf{$\tilde{H}_{Lu}^{(n)}$ }}}
\Text(170,180)[lb]{\small{\textbf{$\tilde{u}^{(0)}$ }}}

\Text(258,125)[lb]{\small{\textbf{$H_u^{(0)}$ }}}
\Text(311,134)[lb]{\small{\textbf{$\tilde{u}^{(0)}$ }}}
\Text(343,65)[lb]{\small{\textbf{$\tilde{Q}^{(0)}$ }}}
\Text(343,134)[lb]{\small{\textbf{$\tilde{u}_L^{(n)}/\tilde{u}_R^{(n)}$ }}}
\Text(290,165)[lb]{\small{\textbf{${A}_{\mu}^{(n)}/{\phi}^{(n)}$ }}}
\Text(363,180)[lb]{\small{\textbf{$\tilde{u}^{(0)}$ }}}

\Text(65,-10)[lb]{\small{\textbf{$H_u^{(0)}$ }}}
\Text(145,20)[lb]{\small{\textbf{$\tilde{Q}_L^{(n)}$ }}}
\Text(145,-65)[lb]{\small{\textbf{$\tilde{u}_L^{(n)}$ }}}
\Text(195,30)[lb]{\small{\textbf{$\tilde{Q}^{(0)}$ }}}
\Text(195,-70)[lb]{\small{\textbf{$\tilde{u}^{(0)}$ }}}

\Text(258,-10)[lb]{\small{\textbf{$H_u^{(0)}$ }}}
\Text(310,-0)[lb]{\small{\textbf{${Q}_L^{(n)}$ }}}
\Text(310,-40)[lb]{\small{\textbf{${u}_L^{(n)}$ }}}
\Text(355,-10)[lb]{\small{\textbf{$\lambda^{(n)}$ }}}
\Text(378,30)[lb]{\small{\textbf{$\tilde{Q}^{(0)}$ }}}
\Text(378,-70)[lb]{\small{\textbf{$\tilde{u}^{(0)}$ }}}
\DashLine(60,120)(120,120){5} \DashLine(120,120)(142,141){5}
\ArrowLine(142,141)(162,160) \DashLine(162,160)(182,180){5}
\ArrowArc(152,150.5)(15,42.52,223.29)
\DashLine(120,120)(181,60){5} \DashLine(60,-15)(120,-15){5}
\DashArrowArc(150,-15)(30,0,360){5} \DashLine(180,-15)(225,45){5}
\DashLine(180,-15)(225,-75){5} \DashLine(253,120)(315,120){5}
\DashLine(315,120)(375,180){5} \DashLine(315,120)(378,59){5}
\PhotonArc(345.31,151.61)(15,42.52,228.29){-5}{8.5}
\DashLine(253,-15)(315,-15){5} \ArrowLine(315,-15)(346,15)
\ArrowLine(346,-45)(315,-15) \ArrowLine(346,15)(347,-45)
\DashLine(346,15)(375,45){5} \DashLine(346,-45)(375,-75){5}
\Gluon(346,15)(346,-45){5}{6}
  \end{picture}
}
\end{center}
\end{center}
\caption{ \small{\em  Feynman diagrams showing the KK loop contribution to the
  evolution of a trilinear scalar coupling.  The diagrams contributing to
  other trilinear couplings may be drawn analogously.}}
 \label{trilinear}
 \end{figure}
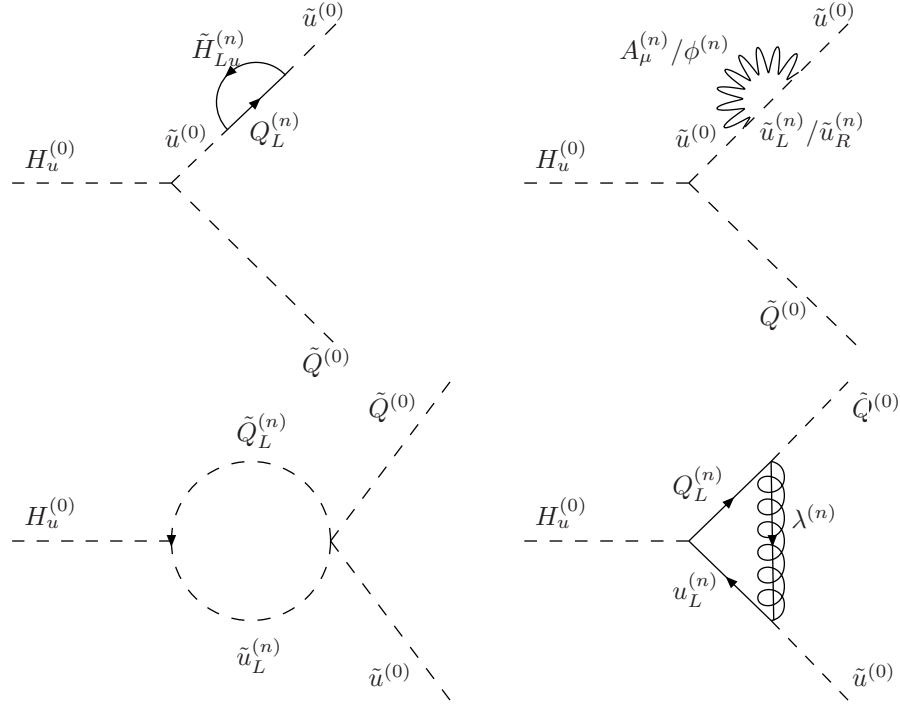
 \end{center}
\normalsize

\subsection{Scalar masses} \label{scalarmasses}
We make three observations regarding the KK contributions to the evolution of
scalar masses (see Fig.~\ref{3higg}):  
\begin{enumerate}

\item The two diagrams in the lower row of Fig.~\ref{3higg} depend on the
  Yukawa couplings. Hence, they are important only for the third generation
  matter fields. 

\item Recall that in the evolution of the Yukawa couplings the KK
  contributions from the gauge field $A_\mu$ and the complex scalar $\phi$
  exactly cancelled thanks to the bulk $N=2$ non-renormalization. However,
  their fermionic superpartners contribute to the scalar mass evolution and
  those contributions add up instead of canceling out. This happens because
  these contributions yield gaugino masses which are $N=1$ supersymmetry
  breaking parameters and hence the non-renormalization theorem ceases to be
  applicable.

\item Each KK state in the two diagrams in the top row of Fig.~\ref{3higg}
  contributes twice that of the SM because of the doubling of the fermions
  (this factor of 2 is highlighted in bold-face in Eqs.~(\ref{scalarbeta}) and
  (\ref{higgsbeta}) below, and also marked on the two diagrams in top row in
  Fig.~\ref{3higg})). However, each KK state at the lower row diagrams
  contributes the same as in the SM because the odd fermion modes vanish at
  the brane where Yukawa interaction is confined.

\end{enumerate}

The zero mode contributions to the beta functions can be found in
\cite{Martin:1997ns}. Below we display the KK contributions to the beta
functions only for the third generation scalars in external lines ($S \equiv
{\rm Tr} [Y_i m_{\phi_i}^2]$, where the index $i$ runs over the third
generation matter states which are in the bulk):
\begin{eqnarray}
\label{scalarbeta}
\tilde{\beta}_{u_3} &=& \frac{1}{16\pi^2} \left[ 2 \left( 2y_t^2 \left(
  m_{Hu}^2 + m_{\tilde{Q}_3}^2 + m_{\tilde{u}_3}^2 \right) + 2a_t^2 \right)
  \right.- \left. {\bf 2} \left( \frac{32}{3}g_3^2|M_3|^2 +
  \frac{32}{15}g_1^2|M_1|^2 + \frac{4}{5}g_1^2 S \right) \right], \nonumber
\\ \tilde{\beta}_{d_3} &=& \frac{1}{16\pi^2} \left[ 2\left( 2y_b^2 \left(
  m_{Hd}^2 + m_{\tilde{Q}_3}^2 + m_{\tilde{d}_3}^2 \right) + 2a_b^2 \right)
  \right. - \left. {\bf 2} \left( \frac{32}{3}g_3^2|M_3|^2 +
  \frac{8}{15}g_1^2|M_1|^2 - \frac{2}{5}g_1^2 S \right) \right], \nonumber
\\ \tilde{\beta}_{Q_3} &=& \frac{1}{16\pi^2} \left[ \left(2 y_t^2 \left(
  m_{Hu}^2 + m_{\tilde{Q}_3}^2 + m_{\tilde{u}_3}^2 \right) + 2a_t^2 \right)
  \right. + \left. \left( 2y_b^2 \left( m_{Hd}^2 + m_{\tilde{Q}_3}^2 +
  m_{\tilde{d}_3}^2 \right) + 2a_b^2 \right)\right. \nonumber\\ & &
  \left. - {\bf 2} \left( \frac{32}{3}g_3^2|M_3|^2 + 6g_2^2|M_2|^2 +
  \frac{2}{15}g_1^2|M_1|^2 - \frac{1}{5}g_1^2 S \right) \right],
\\ \tilde{\beta}_{L_3} &=& \frac{1}{16\pi^2} \left[ \left( 2y_{\tau}^2 \left(
  m_{Hd}^2 + m_{\tilde{L}_3}^2 + m_{\tilde{e}_3}^2 \right) + 2a_{\tau}^2
  \right) - {\bf 2} \left( \frac{6}{5}g_1^2|M_1|^2 + \frac{3}{5}g_1^2 S \right)
  \right],\nonumber \\ \tilde{\beta}_{e_3}
&=& \frac{1}{16\pi^2} \left[ 2\left( 2y_{\tau}^2 \left( m_{Hd}^2 +
  m_{\tilde{L}_3}^2 + m_{\tilde{e}_3}^2 \right) + 2a_{\tau}^2 \right) - {\bf 2}
  \left( \frac{24}{5}g_1^2|M_1|^2 - \frac{6}{5}g_1^2 S \right) \right] \, .
\nonumber
\end{eqnarray}

The beta functions for the Higgs scalars are given by 
\begin{eqnarray}
\label{higgsbeta}
\tilde{\beta}_{H_u} &=& \frac{1}{16\pi^2} 
\left[ 3 \left( 2y_t^2 \left( m_{Hu}^2 +
  m_{\tilde{Q}_3}^2 + m_{\tilde{u}_3}^2 \right) + 2a_t^2 \right) \right.-
  \left. {\bf 2} \left( 6g_2^2|M_2|^2 + \frac{6}{5}g_1^2|M_1|^2 -
  \frac{3}{5}g_1^2 S \right) \right] \, , \nonumber \\ 
\tilde{\beta}_{H_d} &=& \frac{1}{16\pi^2}
\left[ 3 \left( 2y_b^2 \left( m_{Hd}^2 + m_{\tilde{Q}_3}^2 + m_{\tilde{d}_3}^2
  \right) + 2a_b^2 \right) \right.+ \left( 2y_{\tau}^2 \left( m_{Hd}^2 +
  m_{\tilde{L}_3}^2 + m_{\tilde{e}_3}^2 \right) + 2a_{\tau}^2 \right)
  \\ & &- \left. {\bf 2} \left( 6g_2^2|M_2|^2 +
  \frac{6}{5}g_1^2|M_1|^2 + \frac{3}{5}g_1^2 S \right) \right] \nonumber \, .
\label{higgsrg}
\end{eqnarray}

\begin{center}
\begin{figure}
\fcolorbox{white}{white}{
  \begin{picture}(469,257) (0,0)
    \SetWidth{0.5}
\Text(85,112)[lb]{\small{\textbf{$H_u^{(0)}$ }}}
\Text(238,112)[lb]{\small{\textbf{$H_u^{(0)}$ }}}
\Text(135,85)[lb]{\small{\textbf{$\tilde{H}_{Lu}^{(n)},
\tilde{H}_{Ru}^{(n)}$}}}
\Text(150,150)[lb]{\small{\textbf{$\lambda^{(n)}, \psi^{(n)}$ }}}
\DashLine(74,107)(136,107){5} \ArrowLine(136,107)(198,107)
\Text(210,130)[lb]{\small{\textbf{$\times~2$ }}}
\DashLine(198,107)(258,107){5} \ArrowArc(167,107)(30,0,180)
\GlueArc(167,107)(30,0,180){5}{6.87}
\Text(300,112)[lb]{\small{\textbf{$H_u^{(0)}$ }}}
\Text(445,112)[lb]{\small{\textbf{$H_u^{(0)}$ }}}
\Text(345,130)[lb]{\small{\textbf{}}}
\Text(420,150)[lb]{\small{\textbf{$\times~2$ }}}
\Text(310,170)[lb]{\small{\textbf{$\phi_L^{(n)},\phi_R^{(n)}$ }}}
\DashLine(285,107)(468,107){5} \Photon(376,107)(376,153){7.5}{2}
\DashArrowArc(376,173)(20,0,360){5} \Text(85,7)[lb]
{\small{\textbf{$H_u^{(0)}$ }}}
\Text(238,7)[lb]{\small{\textbf{$H_u^{(0)}$ }}}
\Text(110,50)[lb]{\small{\textbf{$\tilde{Q}_L^{(n)},\tilde{u}_L^{(n)}$ }}}
\DashLine(75,2)(255,2){5} \DashArrowArc(155,30)(30,-68,68){5}
\DashArrowArc(175,30)(30,112,248){5}
\Text(300,7)[lb]{\small{\textbf{$H_u^{(0)}$ }}}
\Text(445,7)[lb]{\small{\textbf{$H_u^{(0)}$ }}}
\Text(373,-13)[lb]{\small{\textbf{${Q}_L^{(n)}$ }}}
\Text(373,40)[lb]{\small{\textbf{${u}_L^{(n)}$ }}}
\DashLine(285,2)(347,2){5} \ArrowLine(347,2)(407,2)
\DashLine(407,2)(469,2){5} \ArrowArc(377,2)(30,0,180)
  \end{picture}
}
\caption{\small {\em Feynman diagrams showing the KK contributions to the
    running of the up-type Higgs mass. Diagrams contributing to the evolution
    of the third generation soft scalar masses may be drawn analogously. The
    multiplication factor of 2 has been explained in the text - see point 3
    under section \ref{scalarmasses}.}}
\label{3higg}
\end{figure}
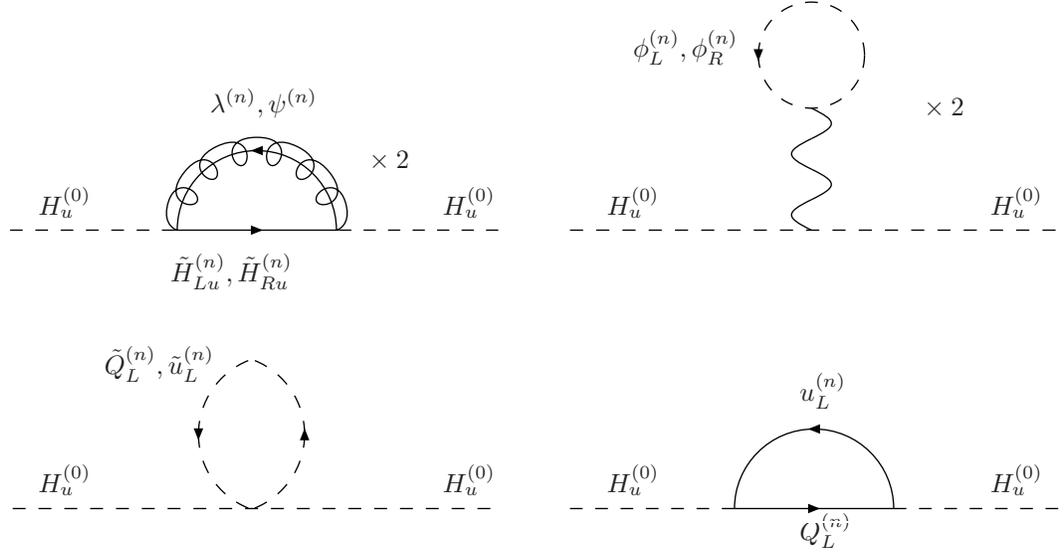
\end{center}
\normalsize
Note that for the first and second generation soft scalar masses only the
second diagram in the top row of Fig.~\ref{3higg} would contribute.

\section{Special numerical features of  RG running in  5d scenario}
\label{numerical}
In this section, we highlight the special features of RG evolution in the 5d
scenario. We also compare and contrast them with the 4d features.  For all our
numerical estimates we have fixed $1/R = 1$ TeV.

\subsection{The Gauge and Yukawa couplings}
The power law running of the gauge and Yukawa couplings has been discussed in
\cite{Dienes:1998vg,blitz} for the non-supersymmetric scenario and in
\cite{Dienes:1998vg} for the supersymmetric case. As far as the Higgs
multiplets are concerned, there is a crucial difference between our model and
that considered in \cite{Dienes:1998vg}.  In our scenario there are separate
up- and down-type Higgs hypermultiplets - see Eq.~(\ref{huhd-hyper}).  Inside
each hypermultiplet only the left column with label ($L$) is $Z_2$ even and
its scalar zero mode receives a vev, whereas the right column with label ($R$)
is projected out.  In other words, the hypermultiplet $H_u$ contains the vev
$v_u$ and, similarly, $H_d$ contains $v_d$.  On the other hand,
\cite{Dienes:1998vg} contains a single hypermultiplet, each column of which
has a zero mode, one to be identified with the up-type chiral multiplet which
contains the vev $v_u$, and the other to be identified with the down-type
containing $v_d$. While our approach constitutes a straightforward
generalization of $H_u$ and $H_d$ from chiral multiplets to hypermultiplets,
the choice made in \cite{Dienes:1998vg} requires non-trivial boundary
conditions.  These two different assumptions lead to significant numerical
differences.  In our approach, the gauge couplings converge to one another but
actually do not meet at a single point, while in \cite{Dienes:1998vg} the
gauge couplings do meet at a point. The difference in the number of KK scalar
excitations makes the difference between the two approaches. 

Indeed, both gauge and Yukawa couplings exhibit power law running due to
summation over the KK states as one crosses the energy thresholds.  As we have
mentioned in section \ref{5dmssm}, keeping the first two matter generations
confined at the brane ensures that the couplings remain perturbative even with
$R^{-1}$ as low as 1 TeV.  Starting from their LEP-measured values at the weak
scale, as we extrapolate the gauge couplings using the KK beta functions we
observe that the three couplings approach very close to one another near 32
TeV, but they do not actually meet at any point, as mentioned in the previous
paragraph.

A crucial point of immense numerical significance is that on account of the
special $N=2$ bulk non-renormalization, the third generation matter
hypermultiplet kept in the bulk does not receive any wave-function
renormalization from the gauge hypermultiplet, which we have illustrated below
Eq.~(\ref{nonrenorm}).  As an important consequence of this the Yukawa
coupling blows up to large (non-perturbative) values around $\Lambda \sim$ 18
TeV, which we therefore take to be the effective cutoff of our theory.

\subsection{The gaugino and scalar masses}
We assume that at the highest scale $\Lambda = 18$ TeV of our theory,
i.e. just before the Yukawa couplings blow up, all scalar masses unify to
$m_0$ and all gaugino masses to $M_{1/2}$. Our high scale parameters are then
$m_0, M_{1/2}, \mbox{sgn}(\mu)$ and $\tan\beta \equiv v_u/v_d$.

\begin{center}
  \begin{figure}[t]
\centering \includegraphics[width=.45\textwidth,angle
  =270,keepaspectratio]{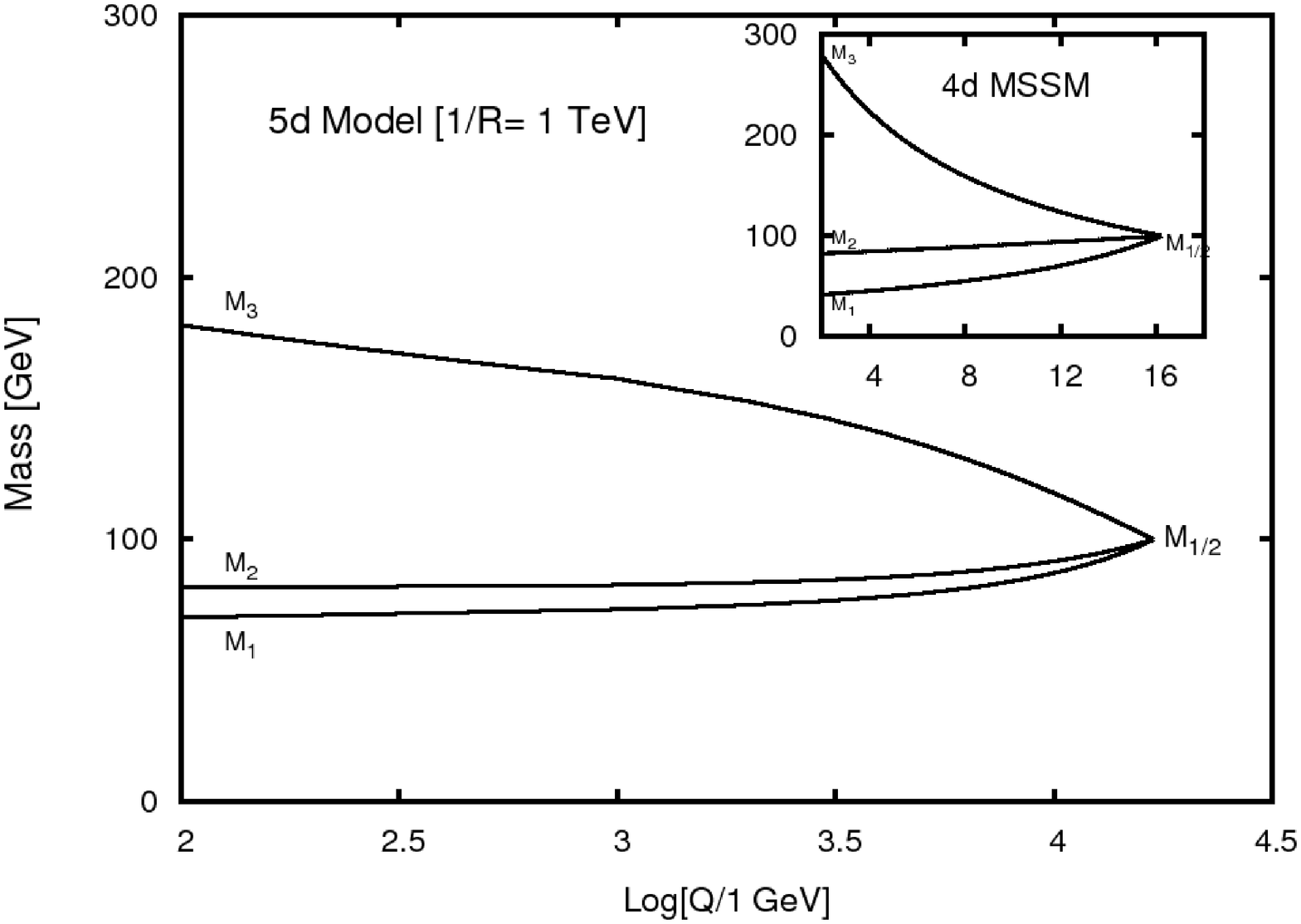}

\caption {\em \small RG running of the gaugino masses.}
\label{gauge}
\end{figure}
\end{center}

The gaugino mass running is governed by the evolution of gauge
couplings. Since gauge couplings {\em nearly} meet around 32 TeV, the gaugino
masses tend to converge also at that scale.  But in the present context, as
mentioned before, we forced the gaugino masses to unify at 18 TeV.  Recall
that in 5d the running is short but fast (power law), but in 4d it is long and
slow (logarithmic). This leads to a general expectation that starting from a
given high scale value, the low scale predictions would be similar in 4d and
5d. But since we forcibly unified the gaugino masses in our set-up, earlier
than otherwise expected, we obtain a somewhat different set of low scale
values. The gaugino mass scaling in 5d is shown in Fig.~\ref{gauge}, while in
the inset is displayed the 4d running.  A rough comparison of the weak scale
ratios of the three gaugino masses is the following:
\begin{eqnarray}
 M_1,M_2,M_3 &\sim& (0.4,0.8,3.0)\times M_{1/2}~ \mbox{(in 4d)}\, , \nonumber
  \\ M_1,M_2,M_3 &\sim& (0.7,0.8,2.0)\times M_{1/2}~
  \mbox{(in 5d)} \, .
\label{gauginom}
 \end{eqnarray}
 If $R$-parity remains conserved, the lightest neutralino remains the lightest
 supersymmetric particle (LSP), only that its mass is heavier than what is
 expected in the standard 4d scenario - see Eq.~(\ref{gauginom}).

\begin{center}
 \begin{figure}[h]
\centering
 \includegraphics[width=.45\textwidth,angle =270,keepaspectratio]{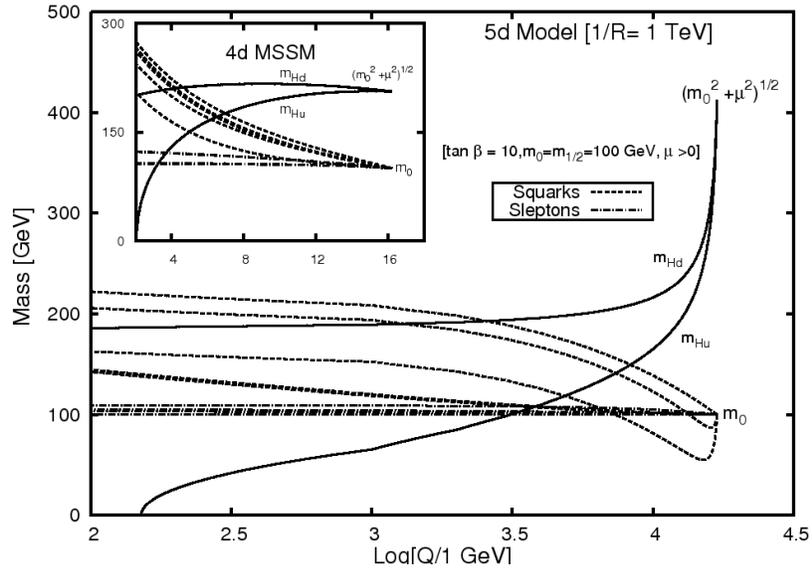}
 \caption{\em \small RG running of the scalar masses and radiative electroweak
   symmetry breaking.}
\label{rewsb}
\end{figure}
\end{center}

 Fig.~\ref{rewsb} shows the running of the soft scalar masses.  The large
 top quark Yukawa coupling continues to play a crucial r\^ole as in 4d. A
 rough comparison of the weak scale predictions in 4d and 5d is:
\begin{eqnarray}
m_{\tilde{Q}_3}^2&\sim& m_0^2 + 5.5 M_{{1}/{2}}^2~ \mbox{(in 4d)} \, , 
\nonumber\\  m_{\tilde{Q}_3}^2&\sim& m_0^2 +
3.5 M_{{1}/{2}}^2~ \mbox{(in 5d).}
\label{scalarm}
\end{eqnarray}
Even for the brane localized scalars, the 5d model predicts slightly higher
weak scale masses compared to 4d\footnote{When the $N=1$ supersymmetry
  breaking is induced by brane-localized spurion vev, the structure of the
  third generation soft masses may, in principle, be different from what we
  assumed due to KK mixing. For calculational simplicity we have ignored KK
  mixing which, we believe, would not affect our main qualitative
  features.}. The relatively {\em compressed} spectrum in 5d, as expressed
through Eqs.~(\ref{gauginom}) and (\ref{scalarm}), is indicative of a lesser
fine-tuning compared to the conventional MSSM.

During power law running we ensure that radiative breaking of electroweak
symmetry does happen at the desired scale\footnote{Radiative electroweak
  symmetry breaking has been discussed in the context of some specific
  realization of supersymmetry breaking in an orbifold
  \cite{Barbieri:2000vh,ArkaniHamed:2001mi,Kobayashi:1998ye}.}. Just like in
4d, only $m_{H_u}^2$ turns negative while all other scalars remain
positive. Again, the large top quark Yukawa coupling drives this phase
transition. A point to note is that {\em unless} we keep the third generation
matter in the bulk, electroweak symmetry would never break radiatively in our
class of models.  Furthermore, we had to take in 5d a factor 1 to 3 larger
(than 4d) value of $\mu$ at the cutoff scale.

\section{The $m_0 -M_{1/2}$ parameter space}
\label{m0mhalf}

\subsection{Numerical procedure} 
For our numerical estimates we go through the following steps:
\begin{enumerate}
\item We scan $m_0$ and $M_{1/2}$ through the range $[0.1-1.0]/R$. We fix
  $\tan\beta = 10$ and take both positive and negative values of $\mu$. Our
  cutoff is 18 TeV, which is the scale where the Yukawa couplings blow up to
  non-perturbative values. We use one loop RG equations as displayed in
  section \ref{beta}. To obtain the weak scale parameters we follow the
  logarithmic evolution like in the 4d desert, but now using different beta
  functions for different energy intervals separated by successive KK
  thresholds - see Eq.~(\ref{master}). So, operationally, power law running is
  treated by taking {\em many} logs.

\item For each input combination, we perform a consistency check to ensure
  correct electroweak symmetry breaking, and accept only those inputs which
  admit this phenomenon.

\item We then feed the weak scale spectrum into the code {\sf micrOMEGAS}
  \cite{Belanger:2001fz}, and using this software package calculate the dark
  matter density ($\Omega_{\rm DM}$), ${\rm Br}~(b \rightarrow s \gamma)$,
  $\Delta a_\mu = (g-2)_\mu/2$, and $\Delta \rho$. Since we consider $1/R = 1$
  TeV, even the lightest KK states are somewhat heavier than the lighter
  section of the zero mode spectra. As a result, without any significant
  numerical compromise we neglect the direct loop contributions of the virtual
  KK particles\footnote{Although $b\to s \gamma$, $Z \to b\bar{b}$, $\Delta
    \rho$, $B_q$--$\overline{B_q}$ are all finite {\em upto one loop} in UED
    or its supersymmetric version thanks to the KK number conservation at the
    {\em tree} level Lagrangian, in scenarios containing brane-bound fermions
    one encounters a UV sensitivity even at one loop \cite{db}. A sensible UV
    completion is needed to counter this. See also \cite{Hebecker:2002vm} and
    footnote 2 of this paper.}. In other words, the KK effects feed into the
  calculation of low energy spectra {\em via} power law running, but after
  that we rely on the standard 4d computations encoded in {\sf
    micrOMEGAS}. This approximation is good enough for our purpose.

\item We compare the predictions of the above observables with their
  experimental values/constraints, and translate the information into the
  inclusion/exclusion plots (Figs.~\ref{momh1} and \ref{momh2}) in the $m_0
  -M_{1/2}$ plane. The 4d plots have been reproduced to serve as a guide to
  the eyes for capturing the 5d subtleties. We note that our 4d plots are in
  agreement with the ones in the existing literature, e.g. with
  \cite{Djouadi:2006be}.
\end{enumerate}

\subsection{Comparison between 4d and 5d models}
We highlight only the major differences between the 4d and 5d models that
appear on the $m_0$--$M_{1/2}$ plane.

\begin{enumerate} 
\item We assume that $R$-parity is conserved.  In the 4d scenario the lightest
  neutralino is the most likely candidate for an LSP.  In the 5d model the
  situation is somewhat tricky. Indeed, the 4d LSP is still an LSP here which
  is the zero mode neutralino. Besides, {\em if} the KK parity remains
  conserved, then the $n=1$ mode of photon tower, namely $\gamma_1$, and its
  supersymmetric partner $\tilde{\gamma}_1$ are also stable dark matter
  candidates. However, the KK parity is unlikely to respected by the
  brane-bulk interaction. In our numerical analysis, we have treated the zero
  mode LSP as the dark matter candidate.  

\item We have taken a $3 \sigma$ range of the five year average of WMAP dark
  matter density ($0.087<\Omega_{\rm DM} h^2<0.138$) \cite{Komatsu:2008hk}. We
  raise a caution here that {\em if} KK parity remains conserved and we have
  two {\em more} dark matter candidates, as mentioned above, then the edge of
  the allowed band arising from the lower limit of $\Omega_{\rm DM}$ would be
  further stretched. Note further that in the 5d case there is a slight
  broadening of the WMAP allowed strip compared to 4d. This happens because of
  a combined effect of Eqs.~(\ref{gauginom}) and (\ref{scalarm}) leading to a
  reduced sensitivity to $M_{1/2}$ variation. 

\item The region where the lightest neutralino satisfies the dark matter
  constraints extends to a {\em higher} value of $M_{1/2}$ in 5d compared to
  4d. What matters most in this context is the mass difference between the
  chargino and the lightest neutralino, i.e. $(M_2 - M_1)$.  It is clear from
  Eq.~(\ref{gauginom}) that starting from a given $M_{1/2}$ this difference is
  smaller in 5d than in 4d. This explains the shift of the band to the right
  side.

\item We have taken $2.65 \times 10^{-4} \lesssim {\rm Br}~(b \to s \gamma)
  \lesssim 4.45 \times 10^{-4}$ \cite{Barberio:2008fa}, and $10.6\times
  10^{-10}\lesssim \Delta a_{\mu}^{\rm new} = (g-2)_{\mu}/2 \lesssim
  43.6\times 10^{-10}$ \cite{Carey:2009zz}. There is nothing much to
  distinguish between 4d and 5d from these two observables.

\item We have {\em not} included the {\em direct} loop effects of the virtual
  KK states for any of the weak scale observables. For $R^{-1} = 1$ TeV or
  more, for processes like muon anomalous magnetic moment or $b\to s \gamma$,
  such effects are numerically negligible, but {\em only } for the Higgs mass
  it makes a difference. In Figs.~\ref{momh1} and \ref{momh2}, the entire
  region to the left of the line marked with $m_h = 114$ GeV which looks
  otherwise disfavored will be resurrected if we include the KK loop
  correction to the Higgs mass \cite{Bhattacharyya:2007te}.
  
\end{enumerate}

\section{Conclusions and Outlook}
\label{concl} 

We reiterate once again that supersymmetry and extra dimension are
interestingly intertwined both from theoretical and phenomenological
perspectives. The presence of extra dimensions is an essential part of any
high scale fundamental theory, and supersymmetry is quite often an integral
component of such theories. Supersymmetry, on the other hand, provides a
sensible UV completion of the extra dimensional models. The connection between
supersymmetry and extra dimension is further strengthened by the consideration
that the origin of supersymmetry breaking may reside in extra dimension. In a
general class of scenarios in which TeV-size soft masses are generated at a
relatively low cutoff scale (like 20-30 TeV) with intermediate resonances, one
can find a common ground where our analysis will be applicable.

The logarithmic running in 4d from 100 GeV to $10^{16}$ GeV is replaced in 5d
by fast power law running on a shorter interval from 100 GeV to about 30 TeV
thanks to the KK states. This is a feature of extra dimension. What is
technically/operationally special about 5d supersymmetry is that it provides a
special $N=2$ non-renormalization that forces us to consider an early cutoff
($\sim$ 18 TeV).

The constraints in the $m_0$--$M_{1/2}$ plane have been placed for the {\em
  first time} in this paper. The ratio $M_1/M_{1/2}$ is higher in 5d compared
to 4d. For this reason the {\em allowed} region in the 5d plot extends to {\em
  larger} values of $M_{1/2}$ compared to the 4d plot. 

It should be noted that $R^{-1}$ could in principle be much smaller, e.g. a
few hundred GeV, than our reference value of 1 TeV. In fact, the first KK
excitations of SM particles could very well be lighter than zero mode
superpartners.  It is neither our intention, nor it is possible within the
scope of the present paper, to distinguish the two different options.

Two issues require further studies beyond the scope of the present paper: (i)
Besides the lightest neutralino (the usual 4d LSP), {\em if} KK parity remains
conserved, there may be two other candidates of dark matter in this model. One
is $\gamma_1$, the $n=1$ level photon, and the other is its superpartner
$\tilde{\gamma}_1$.  Although brane-bulk interface is likely to frustrate KK
parity conservation, yet for the sake of completeness one should study the
{\em combined effects} of all three dark matter candidates.  (ii) It will also
be interesting to revisit the lower limit on $R^{-1}$ in a {\em
  supersymmetric} scenario, which we suspect would be relaxed.

\vskip 5pt

\noindent {\bf{Acknowledgments:}}~ We thank E.~Dudas, S.~Pokorski and
A.~Raychaudhuri for making valuable comments related to our work. GB
acknowledges hospitality of the Physics Department, University of Warsaw,
during the final phase of the work.  GB's work has been supported in part by
the project No.~2007/37/9/BRNS of BRNS (DAE), India.  TSR thanks HRI,
Allahabad, for hospitality during the early phase of this work. TSR also
acknowledges the S.P. Mukherjee fellowship of CSIR, India.



\newpage

\newpage
\centering{
\begin{figure*}
\begin{center}
\includegraphics[width=0.5\textwidth,angle=270,keepaspectratio]
{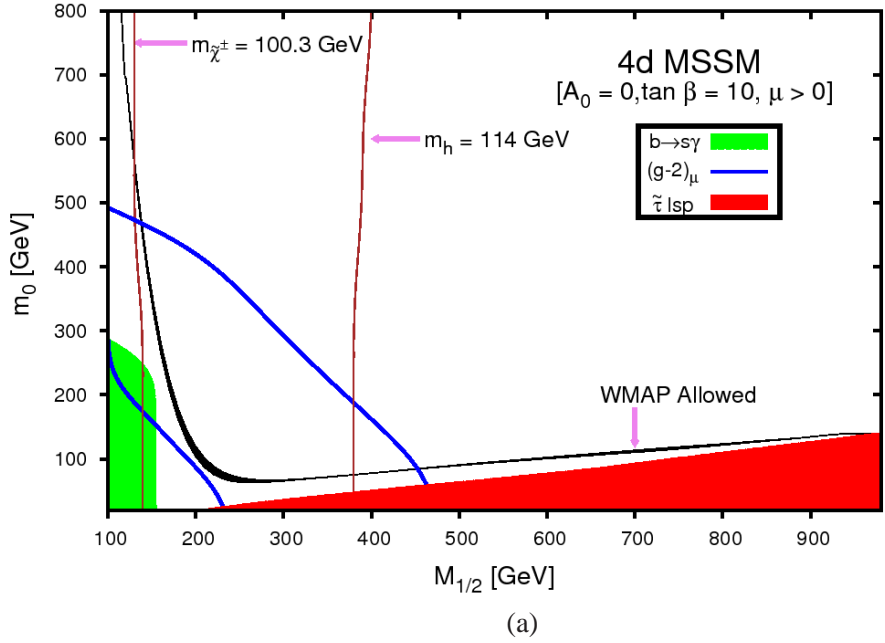}

\centering{\hspace{16mm} (a)}

\includegraphics[width=0.5\textwidth,angle=270,keepaspectratio]
{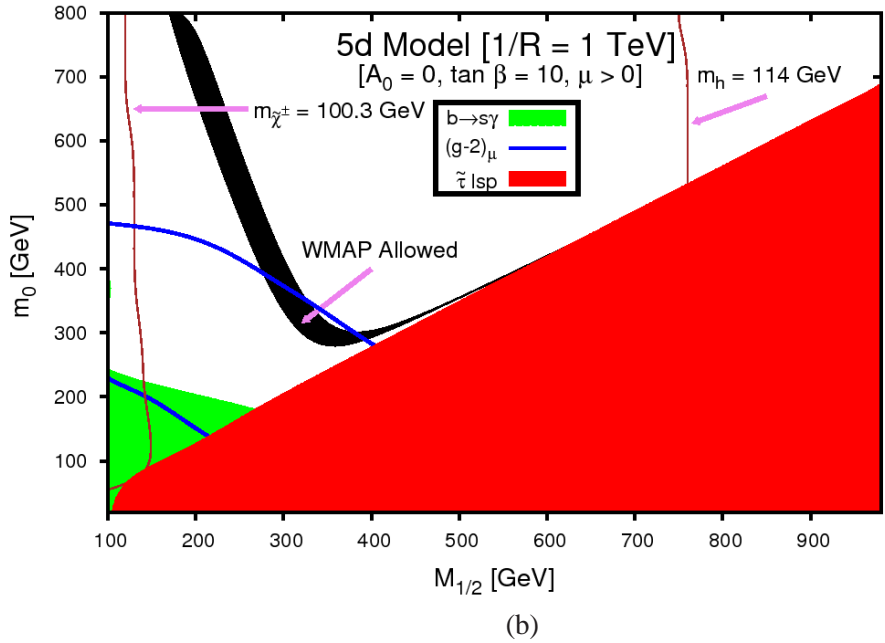}

\centering{\hspace{16mm} (b)}

\end{center}
\caption[]{\em \small The $m_0-M_{1/2}$ parameter space for $\mu>0$. The 4d
  plot is also shown to guide the eye. We keep $tan \beta =10$ for all the
  plots. The region ruled out by ${\rm Br} (b \rightarrow s \gamma)$ is shaded
  in light green (lightest shade) ,the $\tilde{\tau}$ LSP region is shaded in
  red (darker shade) and the region favored by $(g-2)_{\mu}$ is the region
  between the two blue (darkest shade) lines. The WMAP allowed region where
  $.087<\Omega_{DM} h^2<.138$ is shaded in black. We also show the contours
  for $m_h =114 GeV$ and $m_{\tilde{\chi}^{\pm}} =103.3 GeV$, the region to
  the left of these lines are ruled out by LEP exclusion limits. For the 5d
  models, the Higgs contour shown does not include the virtual KK
  contribution.}
\label{momh1}
\end{figure*}
\newpage
\begin{figure*}
\hspace{7mm}
\begin{center}
\includegraphics[width=0.5\textwidth,angle=270,keepaspectratio]
{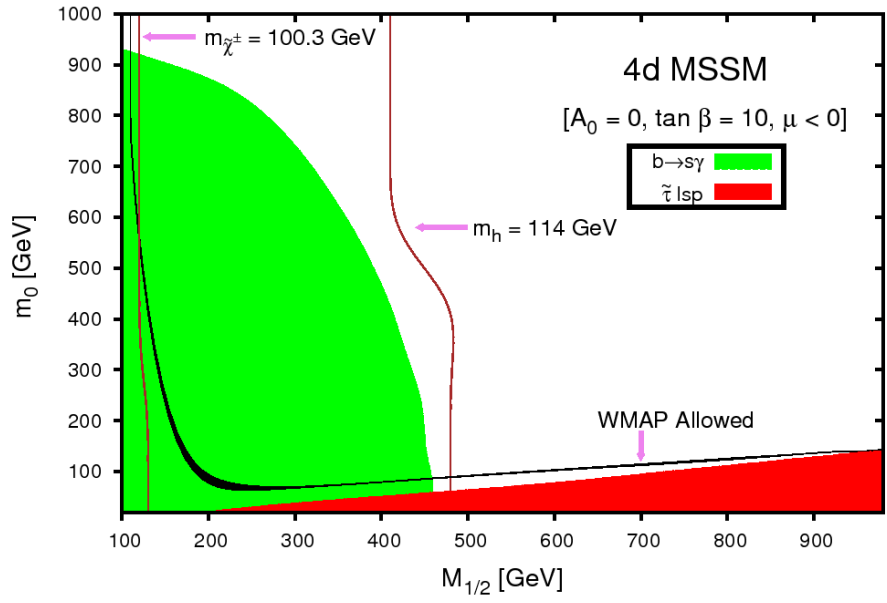}

\centering{\hspace{16mm} (a)}

\includegraphics[width=0.5\textwidth,angle=270,keepaspectratio]
{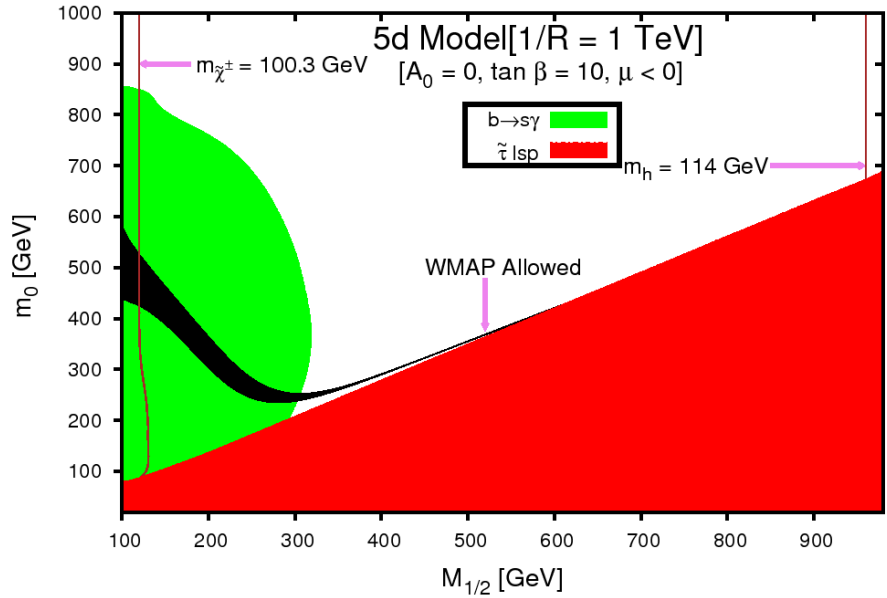}

\centering{\hspace{16mm} (b)}


\end{center}

\caption{\em \small Same as in Fig.~\ref{momh1} but for $\mu<0$. The
          entire region is now disfavored by $(g-2)_{\mu}$. }
\label{momh2}
\end{figure*}
}

\end{document}